\documentclass{article}
\usepackage{spconf,amsmath,xcolor,graphicx,booktabs}

\newcommand{\oldmindcf}{$\mathrm{DCF_{0.01}^{min}}$}
\newcommand{\sreprmmin}{$\mathrm{C_{Prm}^{min}}$}

\title{How to Improve Your Speaker Embeddings Extractor in Generic Toolkits}

\name{Hossein Zeinali$^{1}$, Luk\'{a}\v{s} Burget$^{1}$, Johan Rohdin$^{1}$, Themos Stafylakis$^{2}$, Jan ``Honza'' \v{C}ernock\'{y}$^{1}$}

\address{$^{1}$\:Brno University of Technology, Speech@FIT and IT4I Center of Excellence, Czech Republic \\
$^{2}$\:Omilia - Conversational Intelligence, Athens, Greece }

\begin{document}
\ninept
\maketitle
\begin{abstract}
Recently, speaker embeddings extracted with deep neural networks became the state-of-the-art method for speaker verification. In this paper we aim to facilitate its implementation on a more generic toolkit than Kaldi, which we anticipate to enable further improvements on the method. We examine several tricks in training, such as the effects of normalizing input features and pooled statistics, different methods for preventing overfitting as well as alternative non-linearities that can be used instead of Rectifier Linear Units. In addition, we investigate the difference in performance between TDNN and CNN, and between two types of attention mechanism. Experimental results on Speaker in the Wild, SRE 2016 and SRE 2018 datasets demonstrate the effectiveness of the proposed implementation.
\end{abstract}
\section{Introduction}
\label{sec:intro}

For several years, i-vector representation of a variable length speech signal alongside with Probabilistic Linear Discriminant Analysis (PLDA) has been the state-of-the-art in text-independent speaker verification (TI-SV)~\cite{dehak2011front, prince2007probabilistic}, yielding very good results in other tasks too, such as language identification~\cite{dehak2011language}, text-dependent SV~\cite{zeinali2017hmm, zeinali2017text} and even in non-speech task such as online signature verification~\cite{zeinali2017online}. In recent years, novel deep learning approaches have emerged which outperform the traditional i-vector/PLDA framework.

Deep learning methods for speaker recognition can be summarized into four categories: (a) methods applied to fixed utterance-level representations (typically i-vectors) such as non-linear mappings and backend classifiers ~\cite{stafylakis2012preliminary,pekhovsky2016autoencoders}, (b) i-vectors with Baum-Welch statistics or frame-level features (e.g. bottleneck) extracted with Deep Neural Networks (DNNs) trained for ASR (i.e. with phonetic recognition units as targets)~\cite{lei2014novel,kenny2014deep,lozano2016analysis}, (c) fully end-to-end DNN approaches, where siamese DNNs learn directly to approximate the posterior probability of two or more utterances belonging to the same speaker~\cite{snyder2016deep}, and (d) semi end-to-end approaches, where DNNs with either a closed-set speaker identification architecture (using a softmax over a large number of training speakers) or with a siamese architecture are trained, and utterance-level representations (embeddings) are extracted and fed to a trainable back-end classifier (typically PLDA)~\cite{snyder2018x,nagrani2017voxceleb}. To the best of our knowledge, the performance of the latter category is the current state-of-the-art in most (if not all) speaker recognition benchmarks~\cite{snyder2018x}.

In this paper, we demonstrate how to train a speaker embedding system in a general-purpose deep learning framework and attain comparable (or even better) performance compared to the original Kaldi version~\cite{snyder2018x}. Developing new ideas and combining other proposed method with the x-vector topology is easier in such toolkits, and this is the main motivation for sharing our experience with other researchers. Several papers have been published to show how to train speaker embedding systems in terms of different data augmentation methods and also the amount of required training data~\cite{mclaren2018train, novotny2018use}, but the aim of this paper is to show how to implement an x-vector topology in Tensorflow toolkit, proposing several tricks to improve the performance of speaker embeddings, and empirically evaluate the effectiveness of each trick.

\vspace{-2mm}

\section{System Setup}
\label{sec:system_setup}

In this paper, we focus on speaker embedding training part of the x-vector pipeline and Kaldi toolkit is used for other parts of the pipeline. Our features are 23-dimensional MFCC features, which are extracted from 25\:ms windows with short time mean normalization. Unvoiced frames are eliminated using an Energy based VAD. For creating training archives\footnote{In Kaldi, the network training examples are split to several files which called archive.} for Tensorflow, we use our implementation which produces pretty similar archives like Kaldi except we save minibatches in numpy arrays which saved to tar files. For a fair comparison, all configuration and number of training archives are the same for both Kaldi and Tensorflow and also same Kaldi back-end is used for both implementations.

For training the network we use Adam~\cite{kingma2014adam} optimizer in almost all cases. The initial learning rate is set to 0.001 and linearly reduced to 0.0001. We use 3 epochs for network training. We checked 6 epochs for some systems, but almost all of them overfitted more to the training speakers. In~\cite{mclaren2018train}, it was mentioned 6 epochs is better for Kaldi and our experiments also prove it, but this is not the case for our Tensorflow implementation.

\subsection{Training data and augmentation}

The training data we use in this paper is the list prepared for NIST SRE 2018 close condition and consists of: 1) SREs 4-8 and SRE12, 2) Telephony part of Mixer6, 3) Fisher English, 4) All switchboard data and 5) Voxceleb 1 and 2. For both Voxceleb the concatenated version of each session is used.

The following data augmentation methods are used in this paper. Apart from the four augmentation methods used in~\cite{snyder2018x}, we also include audio compression using ogg and mp3 codecs. Finally, training data consists of 3-fold augmentation that combines {\em clean} data with 2 copies of augmented data, which are selected randomly.

\begin{itemize}
    \item \textbf{Reverberation}: Artificially reverberated data using convolution with simulated RIRs.
    \item \textbf{Babble}: Several speakers are randomly selected from MUSAN~\cite{snyder2015musan} speech and the summation of them is added to the original signal with SNR between 13-20dB.
    \item \textbf{Music}: Adding a random music file from MUSAN to the original signal with random SNR between 5-15dB.
    \item \textbf{Noise}: MUSAN noises are added at one second intervals throughout the recording with random SNR between 0-15dB.
    \item \textbf{Compression}: The original signal is randomly compressed (using ogg or mp3 methods) and it is subsequently converted back to raw format.
\end{itemize}

\subsection{Evaluation data}

We evaluate different networks on three datasets: Speaker in the Wild (SITW) Core-Core condition downsampled to 8\:kHz~\cite{mclaren2016sitw}, the NIST SRE 2016 and the development set of NIST SRE 2018 for Tunisian Arabic (CMN2)\footnote{Note that this results will be replaced with SRE 2018 test set.}. SITW dataset contains recordings extracted from videos in English language and both SRE 2016 and SRE 2018 are conversational telephone speech. We removed the overlapping speakers between SITW and Voxceleb1 from the training data.

\subsection{PLDA backend}

We use a Gaussian PLDA model as back-end classifier. For both SRE 2016 and SRE 2018, the PLDA model is adapted to the unlabeled development data using the unsupervised adaptation of Kaldi and the mean of the unlabeled data to center the x-vectors prior to scoring. For training the PLDA model, a list containing all SREs, Switchboard, Mixer6 and their corresponding augmented versions is used, resulting in about 290 thousand utterances overall. This set is also used for SITW, although this is a suboptimal choice for this set. Moreover, no adaptation technique is used for this set apart from centering the x-vectors using the mean of the SITW development part.

\vspace{-4mm}

\section{Topology and tricks}

Implementing DNN-based methods and replicating published results is a challenging task. In this case, an additional burden is the fact that the original method (x-vector speaker embedding~\cite{snyder2018x}) is implemented in a custom and perfectly tuned toolkit (Kaldi) with several unclear tricks for achieving such a good performance. Here, we try to keep the overall topology same as the original Kaldi model and we investigate the effect of several tricks for boosting its performance for TI-SV. Table~\ref{tbl.xvector_topo} shows the overall topology, which is very close to the original paper~\cite{snyder2018x} and is used as our baseline.

\begin{table}[t]
  \renewcommand{\arraystretch}{1.0}
  \caption{\label{tbl.xvector_topo} Deep neural network topology for x-vector extraction. Here CNNs are used for second and third frame level layers instead of TDNNs.}
  \vspace{1mm}
  \centering{
    \setlength\tabcolsep{3pt}
    \begin{tabular}{c | c | c}
        \toprule
        \textbf{Layer} & \textbf{Layer context} & \textbf{Kernel $\times$ Input $\times$ Output} \\
        \midrule
        Frame1          & $[t-2, t+2]$ & 5 $\times$ 23  $\times$ 512 \\
        Frame2          & $[t-2, t+2]$ & 5 $\times$ 512 $\times$ 512 \\
        Frame3          & $[t-3, t+3]$ & 7 $\times$ 512 $\times$ 512 \\
        Frame4          & $[t]$        & 1 $\times$ 512 $\times$ 512 \\
        Frame5          & $[t]$        & 1 $\times$ 512 $\times$ 1536 \\
        Stats pooling   & $[1, T]$     & 1536 $\times$ 3072 \\
        Segment1        & --           & 3072 $\times$ 512 \\
        Segment2        & --           & 512  $\times$ 512 \\
        Softmax         & --           & 512  $\times$ N \\
        \bottomrule
    \end{tabular}
    \vspace{-5mm}
  }
\end{table}

In the following, the investigated parts of the network and tricks are discussed.

\subsection{Normalizing input features}

It has been proved that normalizing input features has a positive effect on the performance of deep neural networks. Here, our MFCC features are mean-normalized using sliding window. Therefore, the overall features are not normalized and there is a question whether it is useful to normalizing the input features. For this reason, here two different methods are investigated. In the first one, features are simply normalized before feeding them to the network using mean and standard deviation calculated using a subset of training data. In the second method, a Batch-Normalization (BN) layer is added to the input of the network. In the first method, the normalization parameters are kept fixed during training, while in the second method the normalization parameters are learned by the network.

\subsection{Normalizing pooled statistics}

In the original x-vector topology~\cite{snyder2018x}, all layers are followed by a BN layer except the statistic pooling layer. So, the question here is what happened if a BN layer is also added after statistic pooling layer?

\subsection{Order of non-linearity and BN}

Batch-Normalization (BN) is a useful method and helps training deeper networks with fewer epochs and higher learning rate. In~\cite{ioffe2015batch}, the BN layer is placed before the non-linearity while in the x-vector topology it is placed after the non-linearity~\cite{snyder2018x}. Here, we examine the order of the BN layer and non-linearity to show the difference in performance.

\subsection{Avoiding overfitting using dropouts and L2-regularization}
\label{ssec:dropout}

After evaluating the first x-vector implementation in Tensorflow, we observed overfitting to the training speakers compared to the Kaldi version. Assuming segment-level classification accuracy as the measure for overfitting, our implementation attains about 10\:\% better segment accuracy compared to Kaldi for the same training data (i.e. about 95\:\% compared to 85\:\% respectively) and also the SV performance of the Tensorflow version is inferior to that of the Kaldi version for some cases. We therefore examine several methods to prevent the network from overfitting. 

The first regularization method we examine is dropouts~\cite{srivastava2014dropout}, where we test several dropout probabilities. A second method for preventing from overfitting is L2-regularization (also known as L2 weight decay), which penalizes large values in weights, i.e.
\begin{equation*}
    \mathcal{L}^{'} = \mathcal{L} + \beta \: \frac{1}{2} \: {\left\lVert W \right\rVert}_2^2
\end{equation*}
where the best value for $\beta$ should be found empirically. Here, our aim is to answer several questions: for which layers L2-regularization should be used and how much it should participate in the optimization loss (i.e. the value of $\beta$).

\subsection{Feature augmentation using Gaussian noise}

As mentioned in the introduction, several papers investigate the effects of different data augmentations~\cite{mclaren2018train,novotny2018use}. Here we are going to show the effect of adding Gaussian noise to the features during the training. This augmentation is performed in order to minimize overfitting to the training speakers and has a long history in the literature~\cite{graves2006connectionist,graves2013speech}.

\subsection{Different type of non-linearity}

After adding L2-regularization, we faced sparse x-vector representation due to Rectifier Linear Unit (ReLU) saturation. The problem happened for some dimensions, where the ReLU inputs were always negative and so ReLU layer produces only zero output. After adding L2-regularization, the optimizer decides to change the corresponding weights to zero. As a result, the extracted x-vectors were sparse.

Several alternative non-linearities have been proposed for ReLU, from which we test Leaky-ReLU (LReLU) and Parametric-ReLU (PReLU)~\cite{maas2013rectifier,he2015delving}. In LReLU, instead of having zero slope for the negative side of the non-linearity, a small constant slope is used, while in PReLU the slope for the negative region is a trainable parameter and can vary independently for each dimension (making it more vulnerable to overfitting).

\subsection{Comparison between TDNN and CNN}

In the original x-vector paper~\cite{snyder2018x}, Time Delay Neural Network (TDNN) layer is used in the second and third layers of the network. Here, we investigate the differences between TDNN and Convolutional Neural Network (CNN) in performance and also in training and evaluation efficiency. TDNN is a special case of 1-dimensional CNN where instead of using all frames in the context window (convolution window), some specific frames are used (here the first, middle and last frames of the window).

\subsection{Using two types of attention}
\label{sec:attention}

Attention mechanism for speaker verification has been investigated in recent papers. In~\cite{chowdhury2017attention}, several methods were proposed for using attention in an LSTM-based text-dependent speaker verification. A slightly different strategy for adding attention to the x-vector topology was proposed in~\cite{zhu2018self} while single and multi-head attentions were investigated for TI-SV. Here, we only consider single-head attention in two modes. The first one is the same as~\cite{zhu2018self} while for the second one we doubled the size of last hidden layer before pooling and equally split its dimension into two parts like~\cite{chowdhury2017attention} and use the first part for calculating attention weights (i.e. keys) and the second part for calculating mean and standard deviation statistics (i.e. values) using suggested formulas in~\cite{okabe2018attentive}.

\vspace{-4mm}

\section{Experiments and Results}
\label{sec:results}

In order to draw a reliable conclusion about each trick described in the previous section, we performed several experiments. Reporting results for all of them is not possible, hence we only report the most important ones in Table~\ref{tbl.results} and we summarize the remaining in the text.

The first set of experiments is related to normalizing the input features. Adding a BN layer to the input of the network degrades the performance in most cases, while normalizing features using global mean and variance normalization improves the performance in about half cases. Variance normalization of input features is not important, which is in line with the Kaldi implementation where only mean normalization is applied~\cite{snyder2018x}.

Normalizing statistics using the BN layer has a similar trend as normalizing input features and its results were not consistent in all cases. Adding a BN layer after stats pooling using Adam optimizer slightly improves the performance in some cases. But our experiments with SGD optimizer and normalizing statistics using mean and standard deviation calculated in few initial iterations improves the performance. So, it seems this trick is dependent on which optimizer is used. From here, neither input feature normalization nor statistic normalization was used.

Investigating the order of non-linearity and BN layer showed that using BN immediately after non-linearity yields better performance for speaker embedding, while in other fields like image classification~\cite{he2016deep} and audio scene classification (ASC)~\cite{han2017,weiping2017} usually BN layer is used immediately before non-linearity. Our previous experiments in ASC also confirmed that for 2-dimensional CNN network it is better to used BN before ReLU while for x-vector topology (i.e. 1-dimensional network) it is better to put it after the non-linearity~\cite{zeinali2018convolutional}.

As explained in~\ref{ssec:dropout}, we tried to use dropouts to reduce overfitting to the training speakers. Dropouts were shown to improve generalization for classification task, however, our task is to learn speaker representations. Although we observed improved speaker classification performance on our crossvalidation data, the speaker verification performance with the extracted x-vectors degraded for most of the tested dropout probabilities. Also, in our previous work on x-vector based ASC~\cite{zeinali2018convolutional}, dropout helps the performance. It seems that dropouts are useful for classification tasks but not for learning the utterance embeddings.

\begin{table*}[ht]
      \renewcommand{\arraystretch}{1.0}
      \caption{\label{tbl.results} The comparison results of different systems and implementations. All networks use CNN except for them which explicitly named by TDNN. L2 means applying L2-regularization, Att means using attention mechanism in the network and Noise means adding Gaussian noise during training.}
      \vspace{1mm}
      \centerline
      {
  	    \setlength\tabcolsep{3pt}
        \begin{tabular}{l c c c c c c c c c c c c c c c c}
          \toprule
          & & \multicolumn{2}{c}{\bfseries{SITW$_{core-core}$}} & & \multicolumn{2}{c}{\bfseries{SRE16, All}} & & \multicolumn{2}{c}{\bfseries{SRE16, Tagalog}} & & \multicolumn{2}{c}{\bfseries{SRE16, Cantonese}} & & \multicolumn{2}{c}{\bfseries{SRE18, CMN2}} \\
          \cmidrule{3-4} \cmidrule{6-7} \cmidrule{9-10} \cmidrule{12-13} \cmidrule{15-16}
          System & & EER & \oldmindcf & & EER & \oldmindcf & & EER & \oldmindcf & & EER & \oldmindcf & & EER & \sreprmmin \\
          \midrule
            Kaldi recipe, ReLU, TDNN    & & 6.45 & 0.543 & & 8.84 & 0.604 & & 12.72 & 0.764 & & 5.02 & 0.409 & & 9.16 & 0.578 \\
            Kaldi, ReLU, TDNN           & & 5.03 & 0.482 & & 8.02 & 0.566 & & 11.79 & 0.738 & & 4.38 & 0.383 & & 7.30 & 0.501 \\
            Kaldi, ReLU                 & & 4.98 & 0.479 & & 7.81 & 0.566 & & 11.56 & 0.740 & & 4.18 & 0.357 & & 7.44 & 0.504 \\
          \midrule
            TF, ReLU, TDNN              & & 5.08 & 0.500 & & 7.72 & 0.573 & & 11.47 & 0.743 & & 4.08 & 0.359 & & 7.92 & 0.531 \\
            TF, ReLU                    & & 5.33 & 0.517 & & 7.87 & 0.583 & & 11.62 & 0.756 & & 4.15 & 0.362 & & 7.63 & 0.520 \\
          \midrule
            TF, L2, ReLU                & & 4.84 & 0.471 & & 7.59 & 0.568 & & 11.24 & 0.747 & & 4.02 & 0.355 & & 7.57 & 0.517 \\
            TF, L2, PReLU               & & 4.78 & 0.480 & & 7.39 & 0.563 & & 11.01 & 0.742 & & 3.86 & 0.336 & & 7.89 & 0.515 \\
            TF, L2, LReLU               & & 4.73 & 0.467 & & 7.40 & 0.550 & & 11.08 & 0.722 & & 3.79 & 0.340 & & 7.51 & \textbf{0.485} \\
            TF, L2, LReLU, Att          & & \textbf{4.54} & \textbf{0.448} & & \textbf{7.06} & \textbf{0.539} & & \textbf{10.70} & 0.716 & & \textbf{3.47} & \textbf{0.324} & & 7.42 & 0.517 \\
            TF, L2, LReLU, Att, Noise   & & 4.56 & 0.459 & & 7.20 & 0.543 & & 10.74 & \textbf{0.710} & & 3.66 & 0.349 & & \textbf{6.90} & \textbf{0.485} \\
         \bottomrule
        \end{tabular}
        \vspace{-5mm}
      }
\end{table*}

Table~\ref{tbl.results} reports few results of different systems to better compare the gain attained by each technique. The first section of the table shows the results of Kaldi toolkit. The first row shows the Kaldi original recipe for SRE16 where SITW and SRE18 CMN2 development set were added to it with exactly the same training data. By comparing results of this row with the second row, it is clear that on average about 15\:\% relative improvement can be attained by adding more training data and augmentation (or simply having more training speakers).

The third row of the table shows the results of Kaldi toolkit when CNN layers are used in the second and third layers of the network instead of TDNN. In this case, the performance is quite similar to the TDNN while training CNN version needs about 35\:\% more time and also extracting embedding from the network is about 20\:\% slower.

The second section of Table~\ref{tbl.results} shows the baseline results of our TF implementation. Comparing this results with the Kaldi version results shows that our implementation is comparable with Kaldi, sometimes is better and sometimes worse. Here, again the difference between CNN and TDNN is not too much and they performed almost the same.

In the last section of the table, we report results using different tricks for improving our x-vector system. In the first system, L2-regularization was applied to the CNN network (i.e. sixth row of the table). We investigated several configurations for adding L2-regularization. In the simplest way, L2-regularization was applied to all weights of the network while in the second case, it just added to the segment level of the network (i.e. all layers after pooling). Experimental results have shown that the latter case is better and we just consider this case from now.

As explained before, after adding L2-regularization, we faced with sparse x-vectors. For solving this problem, we first remove L2-regularization of the interested embedding layer and it degraded the performance. We also test a smaller coefficient for $\beta$ for this layer and found it was better. Empirically, $\beta$ was set to 0.00002 for embedding layer and 0.0002 for other weights in the segment level of the network. Comparing the results of fifth and sixth rows of the table shows that this simple technique improves the performance about 6\:\% relatively on average.

Although smaller L2-regularization coefficient has better performance, it did not solve the x-vector sparsity. For solving this, we evaluated two other versions of ReLU and their results are shown in the second and third rows of this section. For LReLU, we just select 0.2 for the slope of the negative part and did not check other values. It is obvious that both non-linearities have better performance than ReLU and LReLU performs slightly better. In theory, PReLU should perform better because learns the slope based on the data but it seems it overfitted more to the training speakers.

The two types of attention mechanism described in Section~\ref{sec:attention} were evaluated in this work and we found the variant with separate activations for calculating attention weights and pooled statistics to perform better. The ninth row of the table shows the result of this configuration. This method improves the verification performance for most of the conditions while it increases the computation cost by about 100\:\% in our case.

In the last row of the table, we report the effect of adding Gaussian noise to the features during training as an additional regularization method. For each feature dimension, zero mean Gaussian noise is added with standard deviation of 0.2 times the standard deviation of that dimension. This augmentation improves performance for few cases.

\vspace{-4mm}

\section{Conclusions}
\label{sec:conclusions}

In this work, we have successfully implemented and trained x-vector extractor using a general-purpose machine learning toolkit, namely Tensorflow. We have tested different configurations and modifications to the x-vector extractor topology. We show that using the tricks an suggestions from this paper a similar or better performance can be obtained as compared to the  well tuned original x-vector implementation from the highly optimized Kaldi toolkit.

We tested different normalizations applied to input features and statistics in the pooling layer, but these experiments did not provide consistent improvements over all evaluation datasets. Similarly, we found dropout regularization ineffective when training our speaker embedding extractor. On the other hand, L2-regularization consistently improves the verification performance across all the evaluation conditions.

Both LReLU and PReLU activation functions have improved the verification performance consistently as compared to standard ReLU non-linearity. LReLU performs slightly better than PReLU, which seems to overfitted more to the training data. Attention mechanism have improved the performance for most conditions while it increased the x-vector extraction time by about 100\:\%. However, for the moment, it is not clear whether this improvement comes from the attention mechanism or from the increased number of parameters in the network. This still needs to be investigate in future.

Like other augmentation methods, adding Gaussian noise to the input features during the training has a positive effects on the performance for some conditions. In our experiments, we filter speakers used for training by a minimum number of utterances available per speaker. Adding more augmentations increases the number of utterances available for individual speakers and, as a result, we include more data from more speakers into our training set. Therefore, in future experiments, we should investigate, whether the improvements obtained from the augmentations do not actually come only from having more speaker in the training data.

We will also investigate other neural network architectures, new topologies and training objectives in our future work on learning speaker representations.

\vspace{-4mm}

\section{Acknowledgment}

The work was supported by Czech Ministry of Education, Youth and Sports from Project No. CZ.02.2.69/0.0/0.0/16\_027/0008371, the National Programme of Sustainability (NPU II) project IT4Innovations excellence in science - LQ1602, the Marie Sklodowska-Curie cofinanced by the South Moravian Region under grant agreement No. 665860, and by Czech Ministry of Interior project No. VI20152020025 "DRAPAK".

\vspace{-4mm}

\bibliographystyle{IEEEbib}
\bibliography{refs}

\end{document}